\documentstyle[prl,aps,epsfig,floats,twocolumn]{revtex}
\newcommand{\elimine}[1]{ }
\renewcommand{\log}{{\rm ln}}
\newcommand{\q}{\quad}
\newcommand{\be}{\begin{equation}}
\newcommand{\ee}{\end{equation}}
\newcommand{\ba}{\begin{array}}
\newcommand{\ea}{\end{array}}

\newcommand{\acg}{\left\{}
\newcommand{\acd}{\right\}}

\newcommand{\usd}{{\frac{1}{2}}}
\newcommand{\usq}{{\frac{1}{4}}}
\newcommand{\Tr}[1]{ { \hbox{Tr}\ {#1}  } }
\newcommand{\uni}{{\hat {\bf 1}}}
\newcommand{\ADh}{{\hat{{A}}}}

\newcommand{\HDh}{{\hat{{H}}}}
\newcommand{\KDh}{{\hat{{K}}}}
\newcommand{\MDh}{{\hat{{M}}}}
\newcommand{\PDh}{{\hat{{P}}}}
\newcommand{\NDh}{{\hat{{N}}}}
\newcommand{\QDh}{{\hat{{Q}}}}
\newcommand{\TDh}{{\hat{{T}}}}
\newcommand{\WDh}{{\hat{{W}}}}
\newcommand{\AIh}{{\hat{\cal A}}}
\newcommand{\DIh}{{\hat{\cal D}}}
\newcommand{\avn}{{\langle\NDh\rangle}}
\newcommand{\avm}{{\langle\MDh\rangle}}
\newcommand{\ave}{{\langle\HDh\rangle}}
\newcommand{\pn}{{p}}
\newcommand{\pb}{{\bar p}}
\newcommand{\epsip}{{\epsilon_\pn}}
\newcommand{\ebcsp}{{{\tilde e}_\pn}}
\newcommand{\epn}{{e_\pn}}
\newcommand{\epb}{{e_\pb}}
\newcommand{\apdn}{{a_\pn}}
\newcommand{\apcn}{{a^\dagger_\pn}}
\newcommand{\apdb}{{a_\pb}}
\newcommand{\apcb}{{a^\dagger_\pb}}
\newcommand{\bpdn}{{b_\pn}}
\newcommand{\bpcn}{{b^\dagger_\pn}}
\newcommand{\bpdb}{{b_\pb}}
\newcommand{\bpcb}{{b^\dagger_\pb}}
\newcommand{\deltp}{{\Delta_\pn}}

\newcommand{\up}{{u_\pn}}
\newcommand{\vp}{{v_\pn}}
\newcommand{\tpn}{{t_\pn}}
\newcommand{\qpn}{{t_\pn^{-1}}}
\newcommand{\tpb}{{t_\pb}}
\newcommand{\qpb}{{t_\pb^{-1}}}
\newcommand{\Gpq}{{G_{pq}}}
\newcommand{\qn}{{q}}
\newcommand{\qb}{{\bar q}}

\newcommand{\ebcsq}{{{\tilde e}_\qn}}

\newcommand{\aqdn}{{a_\qn}}

\newcommand{\aqdb}{{a_\qb}}

\newcommand{\deltq}{{\Delta_\qn}}

\newcommand{\tqn}{{t_\qn}}
\newcommand{\qqn}{{t_\qn^{-1}}}
\newcommand{\tqb}{{t_\qb}}
\newcommand{\qqb}{{t_\qb^{-1}}}
\newcommand{\rz}{{r_0}}
\newcommand{\ku}{{k_1}}
\begin{document}
\draft
\twocolumn[\hsize\textwidth\columnwidth\hsize\csname @twocolumnfalse\endcsname
\title{Temperature Dependence of
Even-Odd Effects in Small Superconducting Systems}
\author{R. Balian$^1$, H. Flocard$^2$, and M. V\'en\'eroni$^2$}
\address{$^1$S.Ph.T, CEA, Saclay, F-91191, Gif sur Yvette Cedex\\
$^2$Division de Physique Th\'eorique, 
Institut de Physique Nucl\'eaire, F-91406 , Orsay Cedex}
\date{January 1998}
\maketitle
\begin{abstract}
Generalized BCS equations which consistently include
the projection on the
particle-number parity are derived from
a systematic variational method. Numerical solutions
are given that are
illustrative of ultra-small metallic grains. 
Compared to the BCS approximation, the 
pairing is slightly enhanced in even
and noticeably inhibited in odd systems.
In the latter case, the gap increases
with $T$ near $T=0$; for a suitably small pairing this can result 
in a reentrance phenomenon. Paramagnetic
effects are evaluated as functions of the 
temperature and the magnetic field.
\end{abstract}  

]                                           
\narrowtext

The qualitative differences between the properties of even and
odd nuclei have long been known, as well as their explanation by
pairing correlations\cite{BMo75}.
Even-odd effects have also been displayed in recent years
in superconducting metallic
islands\cite{THT92} and, during the last
two years, in ultra-small Aluminium grains\cite{RBT95}.
In the nuclear
case, the ground and lowest excited states are successfully
described by the BCS theory and its Hartree-Fock-Bogoliubov (HFB)
generalization\cite{RSc80}. Indeed, at zero temperature, the BCS state is a
superposition of configurations with even particle number only,
which allows the currently used models to discriminate between odd and
even systems. 
However, in the case of hot nuclei (which will be investigated 
intensively by means of the next generation of
$\gamma$-detector arrays) or of small metallic superconducting systems, the
non-zero temperature BCS Ansatz mixes configurations with even and odd
particle numbers that have dissimilar physical properties. This additional
spurious mixing {requires projecting the
thermal BCS state onto subspaces having, at least, a well-defined 
number-parity (if not a well-defined particle number)}.
Several theoretical studies have already been devoted to this
question\cite{ANa92}-\cite{BFV97}.

We rely here on a general variational scheme \cite{GRS83} to
derive an extension of the BCS theory which consistently incorporates
the number-parity ($N$-parity) projection.
The method is devised to optimize the (unnormalized) characteristic
function
$\varphi(\xi)\equiv\log{\rm Tr}\PDh_\eta{\rm e}^{-\beta\KDh}\ADh(\xi)$,
with $\KDh=\HDh-\mu\NDh$ and
$\ADh(\xi)\equiv\exp(-\sum_\gamma\xi_\gamma\QDh_\gamma)$ where
$\HDh$, $\NDh$,
$\QDh_\gamma$ and $\xi_\gamma$ are
the Hamiltonian, the particle-number operator,
the observables of interest
and the associated sources.
The trace ($\rm Tr$) is taken over the full Fock space.
In the unnormalized equilibrium density
operator $\PDh_\eta{\rm e}^{-\beta\KDh}$,
the factor $\PDh_\eta$
 is the projection
\begin{displaymath}
\PDh_\eta={\textstyle}\usd(\uni+\eta{\rm e}^{{\rm i}\pi\NDh})
\end{displaymath}
on the subspace with even ($\eta=+1$) or odd ($\eta=-1$) particle
number.
To determine variationally $\varphi(\xi)$
we introduce \cite{BFV97} the action-like functional
\be\label{D02}\ba{l}
\displaystyle
{\cal I}\acg\DIh(\tau),\AIh(\tau)\acd\equiv
{\rm Tr}\DIh(\beta)\exp(-\sum_\gamma\xi_\gamma\QDh_\gamma)\\
\ \displaystyle-
\int_0^\beta{\rm d}\tau\,{\rm Tr}\AIh(\tau)
\Big(\frac{{\rm d}\DIh(\tau)}{{\rm d}\tau}+
\usd[\KDh\DIh(\tau)+\DIh(\tau)\KDh]\Big)\ ,
\ea\ee
where the trial quantities are the operators $\AIh(\tau)$ and
$\DIh(\tau)$. The density-like operator $\DIh(\tau)$ is meant
to include the projection factor $\PDh_\eta$ and to satisfy
the initial boundary condition $\DIh(0)=\PDh_\eta$.
For unrestricted trial spaces, the stationarity conditions of the
functional (\ref{D02}) yield the (symmetrized) Bloch equation for
$\DIh(\tau)$ and, redundantly, the Bloch analogue of the backward
Heisenberg equation for $\AIh(\tau)$ with the final boundary
condition $\AIh(\beta)=\ADh(\xi)$.
When the trial spaces for $\DIh(\tau)$ and $\AIh(\tau)$ are
restricted, the two equations in general become coupled and {\it
the stationary value of the functional} (\ref{D02}) {\it provides
an optimum for} $\exp\varphi(\xi)$.

We are only concerned in this letter by the evaluation of
(projected) thermodynamic quantities. 
These are given by the limit $\xi_\gamma=0$
(i.e., $\ADh(0)=\uni$); indeed
$\varphi(0)$ is the thermodynamic potential associated
with the $\PDh_\eta$-grand canonical equilibrium in which the
particle number can fluctuate but not the $N$-parity.
In this limit and for the trial
spaces (\ref{D05}) that we shall use, one can check that the
optimization of (\ref{D02}) is equivalent
to the usual minimization of the thermodynamic potential\cite{BFV97}.

For the Hamiltonian $\HDh$ we take, as in \cite{BDR97}, the
schematic BCS form 
\begin{displaymath}\label{D03}
\HDh=\sum_p
\epsip(\apcn\apdn+\apcb\apdb)
-B\,{\hat M}
-\sum_{pq}\Gpq\apcn\apcb\aqdb\aqdn\ ,
\end{displaymath}
in which $\apcn$ and $\apcb$ are the creation operators 
associated with the {paired} single-particle states
$\pn$ and $\pb$.
The magnetic-moment operator
${\hat M}\equiv\sum_p\,(\apcb\apdb-\apcn\apdn)$, coupled with
the magnetic field $B$ (in reduced units),
gives rise to paramagnetic effects;
for rotating deformed nuclei, the term $B{\hat M}$
would be replaced by $\omega{\hat J}_z$
where $\omega$ is the angular velocity and
${\hat J}_z$ the $z$-component of the angular momentum.
The pairing coefficients $\Gpq$ are real, symmetric
and, for simplicity,
the diagonal elements $G_{pp}$ are assumed to be zero. 

As variational Ansatz we take
\be\label{D05}
\DIh(\tau)=\PDh_\eta{\rm e}^{-\tau{\WDh}}\ ,\q\q\q
\AIh(\tau)={\rm e}^{-(\beta-\tau){\WDh}}\ ,
\ee
with, for the trial operator $\WDh$, the standard form 
\be\label{D06}
{\WDh}=h_0-\usd\sum_p(\epn+\epb)+
\sum_p(\epn\bpcn\bpdn+\epb\bpcb\bpdb)\ ,
\ee
where $\bpcn$ and $\bpcb$ are the usual quasi-particle operators
$(\bpdn=\up\apdn+\vp\apcb$,
$\bpdb=\up\apdb-\vp\apcn)$.
In the static problem we are considering here, the quantities
$\up$ and $\vp$ can be taken real and non-negative,
with $u_p^2+v_p^2=1$.
Thus our independent variational parameters
are the real numbers $\epn$, $\epb$, $\vp$ and $h_0$.
We get an explicit expression for the functional (\ref{D02})
by inserting Eqs.~(\ref{D05}-\ref{D06})
and using the generalized Wick theorem\cite{BBr69}
for the two terms generated by the projection $\PDh_\eta$.
New projected BCS equations are obtained by writing the
stationarity conditions for this expression.
The familiar BCS equation for the gap is replaced by
\be\label{D07}
\deltp=\usq\sum_q\Gpq\frac{\deltq}{\ebcsq}(\tqn+\tqb)
\frac{1+\eta\rz(\tpn\tpb\tqn\tqb)^{-1}}{1+\eta\rz(\tpn\tpb)^{-1}}\q,
\ee
where the quantities $\tpn$, $\tpb$, $r_0$, $\ebcsq$,
are defined by
\begin{displaymath}
\label{D09}\displaystyle
\tpn\equiv\tanh({\textstyle\usd}\beta\epn)\,,\
\tpb\equiv\tanh({\textstyle\usd}\beta\epb)\,,\
\rz\equiv\prod_p\,\tpn\tpb\ ,
\end{displaymath}
\be\label{D10}
\ebcsp\equiv\sqrt{(\epsip-\mu)^2+\deltp^2}\q.
\ee
The quasi-particle energies $\epn$ and $\epb$ appearing in the
thermal factors $\tpn$ and $\tpb$ differ from the BCS
expression (\ref{D10}); indeed, $\epn$ satisfies
\be\label{D13}
\epn=\ebcsp+B-\frac{\eta\rz\tpn}{\tpn^2-\eta\rz}
\Big(\frac{x_p(\qpn+\qpb)}{1+\eta\rz(\tpn\tpb)^{-1}} + 2\ku\Big)\ ,
\ee
where $x_p$ and $\ku$ are given by 
\begin{displaymath}\label{D08}
x_p\equiv\usq\sum_q\,\Gpq\frac{\deltp\deltq}{\ebcsp\ebcsq}
(\qqn+\qqb-\tqn-\tqb)\q,
\end{displaymath}
\begin{displaymath}
\label{D14}\ba{l}
\ku\Big[
\displaystyle
1+\eta\rz+\eta\rz\sum_p(\displaystyle
\frac{1-\tpn^{2}}{\tpn^{2}-\eta\rz}
+p\leftrightarrow{\bar p}
)\Big]\equiv\\
\displaystyle\q\q
-\usq\sum_p\, x_p
\Big[
(\qpn-\tpn)\frac{\tpn^2+\eta\rz}{\tpn^2-\eta\rz}+
p\leftrightarrow{\bar p}
\Big]\ .
\ea
\end{displaymath}
The quasi-particle energy $\epb$
is obtained from Eq.(\ref{D13}) by the exchanges
$\tpn\leftrightarrow\tpb$ and $B\leftrightarrow -B$.
Altogether, one has to
solve numerically
the self-consistent Eqs. (\ref{D07}) and (\ref{D13})
for  $\deltp$, $\epn$ and $\epb$.
The coefficients $\up$ and $\vp$ are then given explicitly by
the normalization condition $\up^2+\vp^2=1$ and by the relation
$2\up\vp/(\up^2-\vp^2)=\deltp/(\epsip-\mu)$.
All the quantities $\deltp$, $\epn$, $\epb$, $\ebcsp$, 
$r_0={\Tr{{\rm e}^{{\rm i}\pi\NDh}{\rm e}^{-\beta\WDh}}}/
{\Tr{{\rm e}^{-\beta\WDh}}}$, 
$x_p$, $\ku$,
$\up$, $\vp$ 
depend on the $N$-parity
index $\eta=\pm 1$. The standard BCS equations 
are recovered for $\eta=0$.
In Eq.~(\ref{D07}) the projection effects occur 
indirectly via $\epn$ and $\epb$, and directly through
the last fraction.
The form of this fraction entails : 
(i) a predominant influence of the states
near the Fermi level, (ii) odd (even) projected gaps $\Delta_p$
smaller (larger) than their BCS counterparts, and
(iii) greater deviations from BCS for
the odd than for the even systems.

We further assume, as is often done, that the
twofold degenerate single-particle energies
$\epsilon_p$ are
equally spaced with a
level density $w_{\rm F}$.
We use a cut-off energy $\Lambda/2$, above and below
the Fermi level, such that $w_{\rm F}\Lambda=100$.
The pairing coefficients $\Gpq$ are chosen of the
form ${\tilde G}(1-\delta_{pq})$.
The projected and the BCS gaps depend on $\pn$\ ;
however, the variations of $\Delta_\pn$ with $\pn$
never exceed a few percent in our calculations.
A convenient energy scale is provided by
the BCS gap $\Delta$ associated, for $T=0$ and $B=0$,
with the quasi-particle of lowest energy.
This model, depending on the {values} of
$w_{\rm F}\Delta$, or alternatively of $w_{\rm F}{\tilde G}$,
schematically describes
superconducting islands, heavy nuclei or
ultra-small metallic grains. 
The corresponding variational equations above
have been solved in  Ref.\cite{BFV97} for $B=0$.
The results display
the growing magnitude of the even-odd 
 effects
{as $w_{\rm F}\Delta$ decreases}. In view of recent
experimental advances\cite{RBT95},
we give preference here 
to parameters typical of ultra-small grains.

\begin{figure}[h]
\begin{center}
\leavevmode
\hbox{\epsfxsize=7.8 truecm\epsfbox{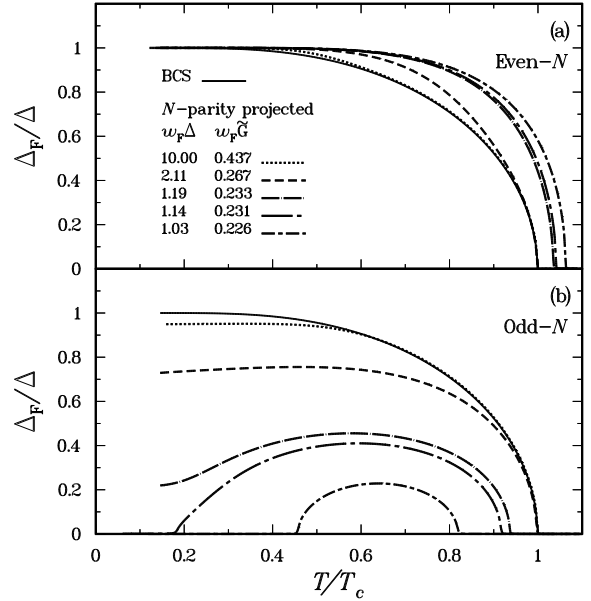}}
\end{center}
\caption[F1]{Even/odd projected gaps.
The gap $\Delta_{\rm F}$ 
is plotted (for $B=0$) as function of $T$
for various values of $w_{\rm F}{\Delta}$, or equivalently
of $w_{\rm F}{\tilde G}$.
The upper part (a) corresponds to the even system $\avn=100$ and the
lower part (b) to the odd system $\avn=101$.}
\end{figure}
For $B=0$ and for various values of
$w_{\rm F}\Delta$, Fig.1 shows the variation with $T/T_{\rm c}$ 
($T_{\rm c}$ denotes the BCS critical
temperature for $B=0$) of the ratio
$\Delta_{\rm F}/\Delta$, where 
$\Delta_{\rm F}$ is the temperature-dependent
(odd or even) gap $\Delta_p$ associated with the quasi-particle
of lowest energy $\epn$.
According to Eq.~(\ref{D07}), both the gap values and the critical
temperatures show that pairing is enhanced (inhibited) by projection
on even (odd) particle number.
In the odd case, near $T=0$, a new phenomenon appears:
{\it the growth of the pairing correlations with temperature}.
This effect can be understood as follows. At $T=0$ the unpaired particle
occupies one of the two states $\epsilon_0=\epsilon_{\bar 0}=\mu$,
precluding them to participate
to the pair components. 
For (small) increasing values of $T$, thermal
excitation induces the migration of this unpaired
particle to higher states,
thus making the strategic states at the Fermi level
available to the formation of pairs.

\begin{figure}[h]
\begin{center}
\leavevmode
\hbox{\epsfxsize=7.truecm\epsfysize=8.truecm\epsfbox{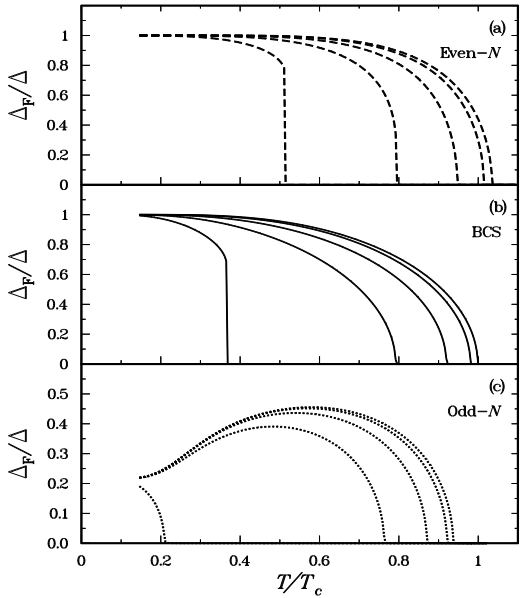}}
\end{center}
\caption[F2]{
Dependence upon $T$
of the gap $\Delta_{\rm F}$
for $w_{\rm F}{\Delta}=1.19$ ($w_{\rm F}{\tilde G}=0.233$)
and for magnetic-field values 
such that $w_{\rm F}B$=0; 0.2; 0.4; 0.6; 0.8.
The three plots correspond to (a) the even-$N$ projection
($\avn=100$), (b)  the BCS theory ($\avn=100$), and
(c) the odd-$N$ projection ($\avn=101$).
Note the scale difference between (c) and (a).}
\end{figure}

For suitable values of $w_{\rm F}\Delta$
(in our example $1. < w_{\rm F}{\Delta} < 1.14$),
this increase of pairing correlations versus $T$ (near $T=0$) can
even generate a {\it reentrance phenomenon} with two transition
temperatures (Fig.~1-b).
In such a case pairing is absent at $T=0$\ ;
as $T$ increases, it switches on at some low critical temperature,
then switches off at the higher 
familiar critical temperature\cite{BFV97,NOTE}.
The two critical temperatures merge for 
$w_{\rm F}{\Delta}\approx 1$,
which corresponds to $w_{\rm F}{\tilde G}\approx 0.225$;
for weaker values there exists
no superconductivity in odd systems,
whatever the temperature. 
In contrast, in even systems, pairing does not disappear until
$w_{\rm F}{\tilde G} < 0.182$.

The comparison of (a) and (b) in Fig.2 shows that,
at all temperatures, the even-$N$ gap resists
magnetic fields better than the BCS gap.
It is also seen on Fig.2 that, at $T=0$,
the gaps $\Delta_{\rm F}$ do not depend on the magnetic fields;
this is because {the field intensities}
are such that $B<\Delta$.
When $B>\Delta$, for the value $w_{\rm F}\Delta\approx 1.19$ of Fig.~2,
pair correlations disappear completely.
By comparing (b) and (c)
one sees that the odd-$N$ gap is significantly smaller than its BCS
counterpart and more easily distroyed by the paramagnetic effects.
Nevertheless, the growth with $T$ of the odd-$N$ gap at
small $T$ persists in the presence of a sufficiently small
magnetic field.

Once the self-consistent equations (\ref{D07}) and (\ref{D13}) are solved,
the {\it projected} average particle number,
magnetization, energy, entropy 
(and hence the other thermodynamic
quantities) can be obtained from the formulae
\begin{displaymath}
\label{D15}\ba{rl}\displaystyle
\displaystyle
\avn=&\displaystyle \sum_p \Big[
1-\frac{\displaystyle \epsip-\mu}{\displaystyle \ebcsp}(\tpn+\tpb)
\frac{\displaystyle 1+\eta\rz(\tpn\tpb)^{-1}}{\displaystyle 2(1+\eta\rz)}
 \Big]\q,\\
\displaystyle
\avm=&\displaystyle \sum_p (\tpn-\tpb)
\frac{\displaystyle 1-\eta\rz(\tpn\tpb)^{-1}}{\displaystyle 2(1+\eta\rz)}\q,\\
\displaystyle
\ave=&\displaystyle \sum_p \Big[
\epsip
-\frac{\displaystyle \epsip(\epsip-\mu)+{\textstyle\usd}\deltp^2}
{\displaystyle \ebcsp}\times
\\&\q\q\q(\tpn+\tpb)
\frac{\displaystyle 1+\eta\rz(\tpn\tpb)^{-1}}{\displaystyle 2(1+\eta\rz)} \Big]
-B\,\avm \ ,\\
S=&\displaystyle S_{\rm BCS}-
\frac{\displaystyle\beta\eta\rz}{\displaystyle 2(1+\eta\rz)}
\sum_p[(\qpn-\tpn)\epn+ p\leftrightarrow{\bar p}]\\
&\q\q\q+\log[\usd(1+\eta\rz)]\q.
\ea\end{displaymath}
These quantities satisfy the fundamental thermodynamic relations,
as warranted on general grounds by the
variational nature of our approximation.
Various limits (low and high temperatures, continuous level density,
large volume) and numerical illustrations are 
worked out in \cite{BFV97} for $B=0$, showing the
qualitative differences between even and odd systems
in physical quantities such as the  specific heat. 

Here we show on Fig.3, for the same values of
$w_{\rm F}{\Delta}$ and of the magnetic fields as in Fig.~2,
the magnetization $\avm$ versus the ratio
$T/T_c$ for the BCS theory,  
for the even ($N=100$) and for the odd ($N=101$)
projected cases. All the curves converge to the Pauli
paramagnetic values $2 w_{\rm F} B$ at high enough temperatures.
Until the $B$-dependent transition temperature is reached, the BCS
magnetization is always larger than the even-$N$ one,
but both behave similarly.
{The behavior at small $T$
of the odd-$N$ curves is markedly different}.
While for $B=0$ the magnetization
is zero as expected, any value of $B$, however small,
lifts the single-particle degeneracy of the spectrum and causes
the zero-temperature value of $\avm$ to take integer values. For
the small magnetic fields considered here ($w_{\rm F} B < 1$),
one has $\avm=1$ for $T=0$. As $T$ grows, 
$\avm$ increases or decreases in odd systems,
depending on the value of $B$,
so as to reach eventually the Pauli
paramagnetic value. 
\begin{figure}[h]
\begin{center}
\leavevmode
\hbox{\epsfxsize=8.truecm\epsfysize=5.truecm\epsfbox{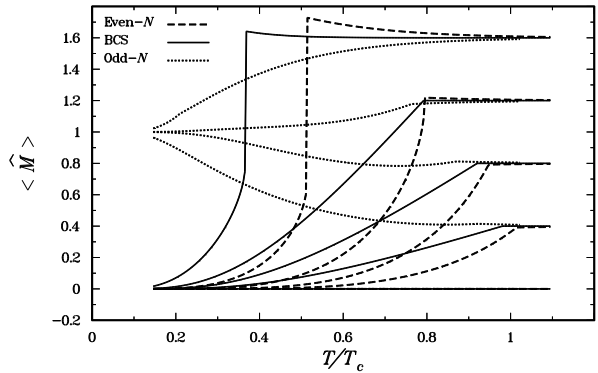}}
\end{center}
\caption[F3]{
The magnetization $\avm$ versus $T/T_c$ for
different values of the magnetic field $B$
and for $w_{\rm F}{\Delta}=1.19$ ($w_{\rm F}{\tilde G}=0.233$).
The dashed lines correspond to the even-$N$ projection
($\avn=100$), the solid lines to BCS ($\avn=100$) and
the dotted lines to the odd-$N$ projection ($\avn=101$).}
\end{figure}

Our curves exhibit a spurious non-analytic
behavior at the critical points.
This is a consequence of the approximation (\ref{D05}) which,
like the BCS one, gives rise to either a broken or an unbroken invariance
with respect to $\NDh$, depending on the temperature.
In the exact treatment of a finite system without sources, 
the order parameter should always be zero 
and {there can be no discontinuity at any temperature}.
However, allowing an order parameter 
while restoring the broken symmetry
may reproduce the rounding-off of the transition\cite{MSD72}.
For instance, in the present problem the difficulty above
may be removed by replacing
$\PDh_\eta$ by the projection
$\PDh_{N_0}=(2\pi)^{-1}\int_0^{2\pi}{\rm d}\theta\,\exp(i\theta[\NDh-N_0])$
on the particle number $N_0$, thus restoring the
invariance broken by $\WDh$.
However, in this case the trial
Ansatz $\DIh(\tau)=\PDh_{N_0}\TDh(\tau)\PDh_{N_0}$,
where $\TDh(\tau)$
is the exponential of the most general
quadratic form in $a^\dagger_i$ and $a_j$,
must include projections on
both sides\cite{BFV97}.
(In contrast, the Ansatz (\ref{D05}) contains a single
projection since $\PDh_\eta$ commutes
with $\WDh$.)
After the substitution of $\PDh_\eta$ by $\PDh_{N_0}$,
which amounts to replacing a two-term sum by a double
integral over  $\theta$,
the functional (\ref{D02}) remains an operational tool
since one can still use Wick's theorem.
The optimization of
(\ref{D02}) then provides a generalization of the
HFB theory which takes properly into account the particle number.
In this scheme, where variation is performed after
projection, the quantities $u_\pn\,v_\pn$ are expected
to be non-zero at any temperature, and to yield smooth transitions.
Likewise, in (cold) deformed nuclei the evolution
of the moment of inertia with respect to the angular momentum
is smoothened by the $\PDh_{N_0}$-projection, in agreement with
the data\cite{GFB94}.

Projected fluctuations and correlations could be
evaluated, along the lines of Sect.4 of \cite{BFV97}, through the
expansion of $\varphi(\xi)$, as given by (\ref{D02}), 
up to second order in the 
sources $\xi_\gamma$. Finally, as proposed in \cite{BVe85}, 
transport properties could be worked out by combining the static variational
principle (\ref{D02}) with a time-dependent one.

One of us (M.V.) gratefully acknowledges the support of the 
Alexander von Humbolt-Stiftung and the 
hospitality of the Physics Department of the
Technische Universit\"at M\"unchen.


\begin{references}
\bibitem{NOTE} The reentrance effect can be inferred
from the Fig.1-a of Ref.\cite{DZG96} by imagining
cuts of the (implicit) surface of this figure by planes
perpendicular to the $d/{\tilde\Delta}$ axis for values 
slightly larger than 0.89, which corresponds to 
$w_{\rm F}\Delta=1.15$ on our Fig.~1-b.
\end{references}
\end{document}